\documentclass[cup6b]{cupbook}
\usepackage{graphicx}
\title[Rome, Italy, 27--30 April 2009]
      {The coming of age of X-ray polarimetry}
\author{}
\date{}
\begin{document}
\pagenumbering{arabic}


\author[Stefano Bianchi et al.]{Stefano Bianchi (Universit\`a degli Studi Roma Tre, Italy) \and Giorgio Matt (Universit\`a degli Studi Roma Tre, Italy) \and Francesco Tamborra (Universit\`a degli Studi Roma Tre, Italy) \and Marco Chiaberge (Space Telescope Science Institute, U.S.A.) \and Matteo Guainazzi (XMM-\textit{Newton} SOC, ESAC, ESA, Spain)  \and Andrea Marinucci (Universit\`a degli Studi Roma Tre, Italy)}

\chapter{The soft X-ray polarization in obscured AGN}

\abstract{The soft X-ray emission in obscured active galactic nuclei (AGN) is dominated by emission lines, produced in a gas photoionized by the nuclear continuum and likely spatially coincident with the optical narrow line region (NLR). However, a fraction of the observed soft X-ray flux appears like a featureless power law continuum. If the continuum underlying the soft X-ray emission lines is due to Thomson scattering of the nuclear radiation, it should be very highly polarized. We calculated the expected amount of polarization assuming a simple conical geometry for the NLR, combining these results with the observed fraction of the reflected continuum in bright obscured AGN.}

\section{Introduction}

The presence of a `soft excess', i.e. soft X-ray emission above the extrapolation of the absorbed nuclear emission, is very common in low resolution spectra of nearby X-ray obscured active galactic nuclei (AGN, Guainazzi, Matt \& Perola, 2005; Turner et al., 1997). It is generally very difficult to discriminate between thermal emission, as expected by gas heated by shocks induced by AGN outflows or episodes of intense star formation, and emission from a gas photoionized by the AGN primary emission. However, the high energy and spatial resolution of XMM-\textit{Newton} and \textit{Chandra} have allowed us to make important progress in the last few years.

\section{The photoionization signatures}

\begin{figure}
\centering
\includegraphics[scale=0.42]{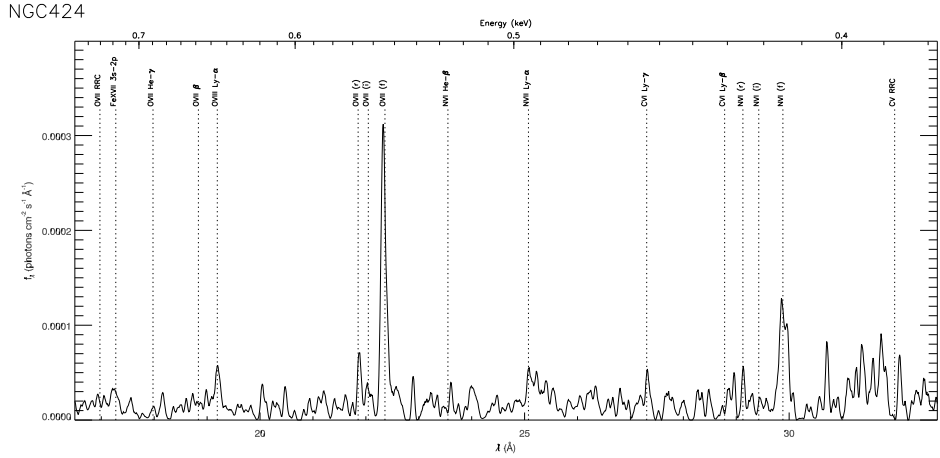}
\caption{RGS spectrum of NGC~424 (a.k.a. Tololo 0109-383). Spectra of the two RGS cameras have been merged and smoothed with a five-channel wide triangular kernel for illustration purposes only. The positions of the line transitions measured in CIELO-AGN are labelled. From Marinucci et al., in preparation.}
\label{tololo}
\end{figure}

The high resolution spectra of the brightest obscured AGN, made available by the gratings aboard \textit{Chandra} and XMM-\textit{Newton}, revealed that the `soft excess' observed in CCD spectra was due to the blending of strong emission lines, mainly from He- and H-like transitions of light metals and L transitions of Fe (see Fig.~\ref{tololo}, e.g. Sako et al., 2000; Sambruna et al., 2001; Kinkhabwala et al., 2002; Brinkman et al., 2002; Schurch et al., 2004; Bianchi et al., 2005). The presence of narrow radiative recombination continua (RRC) features from Carbon and Oxygen, whose width indicates typical plasma temperatures of the order of a few eV, are unequivocal signatures of photoionized spectra (Liedahl \& Paerels, 1996). Moreover, the intensity of higher-order series emission lines, once normalized to the K$\alpha$, are larger than predicted by pure photoionization, and are consistent with an important contribution by photoexcitation (resonant scattering, Band et al., 1990; Matt, 1994; Krolik \& Kriss, 1995; Matt, Brandt \&  Fabian 1996). All these pieces of evidence agree that the observed lines should be produced in a gas photoionized by the AGN, with little contribution from any collisionally ionized plasma.

These results have been confirmed to be common in a large catalog of Seyfert 2 galaxies (CIELO-AGN: Catalog of Ionized Emission Lines in Obscured AGN,  Guainazzi \& Bianchi, 2007), who presented results of a high-resolution soft X-ray (0.2--2~keV) spectroscopic study on a sample of 69 nearby obscured AGN) observed by the Reflection Grating Spectrometer (RGS) on board XMM-\textit{Newton}. Their analysis confirmed the dominance of emission lines over the continuum in the soft X-ray band of these sources, the presence of narrow RRC and the important contribution from higher-order series lines. Therefore, this study allows us to confirm that the results extracted from the detailed study of high-quality spectra of the brightest objects can be extended to the whole population of nearby obscured AGN.

\section{The coincidence with the NLR: the same medium?}

\begin{figure}
\centering
\includegraphics[scale=0.7]{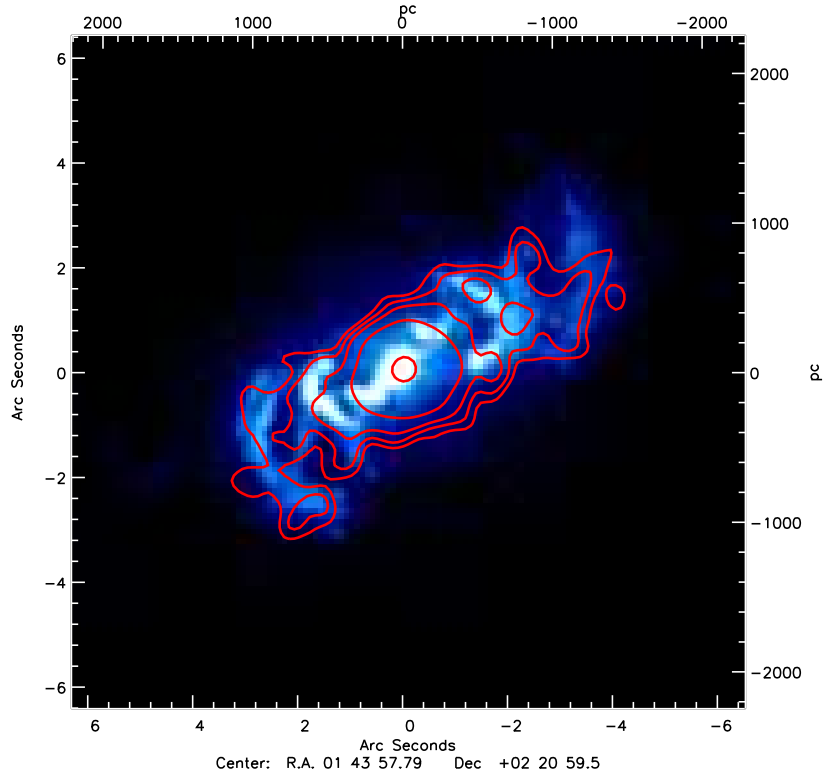}
\caption{Mrk~573: \textit{Chandra} soft X-ray (0.2-2 keV) contours superimposed on the \textit{HST} {O\,\textsc{iii}} image. The contours refers to 0.001, 0.1, 0.2, 0.4 and 0.95 levels with respect to the brightest pixel. North is up, east to the left. From Bianchi et al., in preparation.}
\label{xray2oiiimaps}
\end{figure}

Thanks to the unrivaled spatial resolution of \textit{Chandra}, it has been possible to resolve the soft X-ray emission of Seyfert 2 galaxies, which appears to be strongly correlated with that of the Narrow Line Region (NLR), as mapped by the [{O\,\textsc{iii}}] $\lambda 5007$ \textit{HST} images (see Figure \ref{xray2oiiimaps}, e.g. Young, Wilson \& Shopbell, 2001; Bianchi, Guainazzi \& Chiaberge, 2006; Levenson et al., 2006).

The possibility that the NLR, which is also believed to be gas photoionized by the AGN, is the same material responsible for the soft X-ray emission was investigated in detail by Bianchi, Guainazzi \& Chiaberge (2006). They found that such a simple scenario is tenable.  Moreover, the observed [{O\,\textsc{iii}}] to soft X-ray flux ratio remains fairly constant up to large radii, thus requiring that the density decreases roughly like $r^{-2}$, similarly to what often found for the NLR (e.g. Kraemer et al., 2000; Collins et al., 2005).

\section{Polarization}

High resolution spectroscopy of soft X-ray emission in obscured AGN reveals that it is dominated by strong emission lines. The common spatial coincidence between the soft X-ray emission and the NLR suggests they can also be one and the same medium, photoionized by the central AGN. In this scenario, a fraction of the soft X-ray flux is still expected to be constituted by a featurless continuum, due to Thomson scattering of the primary radiation. This is indeed what is found in low-resolution, CCD spectra, when all the flux detected in emission lines is taken into account.

We are systematically analysing all the brightest obscured AGN observed by the XMM-\textit{Newton} gratings (RGS) and CCD (EPIC), in order to disentangle the fraction of soft X-ray flux in the continuum from that of the emission lines (Tamborra et al., in preparation). First results include the Phoenix Galaxy (60\%, based on \textit{Suzaku} data: Matt et al., 2009), NGC~424 (30\%, XMM-\textit{Newton} RGS and EPIC pn data: Marinucci et al., in preparation), and Mrk~573 (35\%, XMM-\textit{Newton} RGS and \textit{Chandra} ACIS-S: Bianchi et al., in preparation).

\begin{figure}
\centering
\includegraphics[scale=0.7]{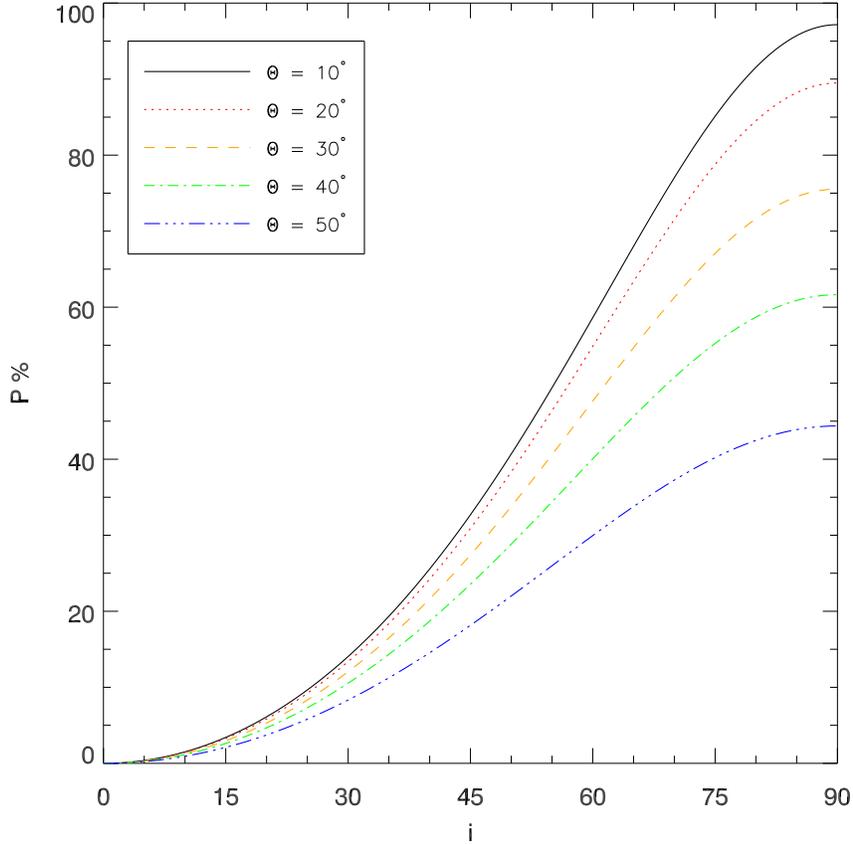}
\caption{Expected polarization degrees for the continuum component of the soft X-ray emission in obscured AGN as a function of the inclination angle $i$ between the line of sight and the axis of symmetry, and for different values of the half-opening angle $\theta$ of the emission cone.}
\label{pollo}
\end{figure}

The fraction of the soft X-ray emission constituted by the continuum should be very highly polarized. Assuming a conical shape and homogeneous density for the soft X-ray emitting region, we estimated the expected polarization degree for this radiation (see Fig. \ref{pollo}). The results shown in the figure have been calculated following Brown \& McLean (1977), who derived the formulae for the flux and polarization of the Thomson scattered radiation in axisymmetric configurations of matter illuminated by a point like source. The formulae are valid in the optically thin case, i.e. in the single scattering approximation. Unpolarized primary radiation is assumed for simplicity. For symmetry reasons, the polarization vector is orthogonal to the axis of the cone. The polarization degree increases with the inclination angle (i.e. the angle between the line of sight and the axis of the cone), because of the well-known dependence of the polarization degree on the scattering angle $\alpha$, i.e. P=(1-cos$^2\alpha$)/(1+cos$^2\alpha$). The polarization degree decreases with the half--opening angle of the cone because of the decreasing level of asymmetry. For typical values of the half--opening angles (30$^{\circ}$--50$^{\circ}$) the polarization degree is very high, up to 45-70\% for an edge--on view. 
 
With the continuum fraction measured for the Phoenix Galaxy, NGC~424, and Mrk~573, these results would correspond to net polarization degrees of around 20\% for the first source (total F$_{0.5-2\,\mathrm{keV}}=2.7\times10^{-13}$ erg cm$^{-2}$ s$^{-1}$), and 10\% for the others (total F$_{0.5-2\,\mathrm{keV}}=1.9\times10^{-13}$ and $2.9\times10^{-13}$ erg cm$^{-2}$ s$^{-1}$, respectively), assuming $i$=50$^{\circ}$ and $\theta$=30$^{\circ}$. These estimates assume that line emission is completely unpolarized, i.e. polarization of resonant scattering lines is not considered.

\section{Conclusions}

The soft X-ray emission in obscured AGN is likely due to continuum and line emission from photoionized
matter in an axisymmetric configuration. Even after dilution by (mostly unpolarized) line emission,
high degrees of polarization are expected. Obscured AGN are therefore among the best candidates for
any polarimeter working in the soft X-ray band.

\end{document}